\documentclass{pasj00}
\def\commenta{$^*$}
\def\commentb{$^\dagger$}
\def\commentc{$^\ddagger$}
\def\commentd{$^\|$}
\def\commente{$^\S$}

\def\vsnetalert#1{VSNET alert circular, #1}

\def\labelspace{}
\def\pasjcitep2sub#1#2{(\authorcite{#1} \yearcite{#1},\yearcite{#2})}

\DeclareAbbreviation\an{Astron. Nachr.}
\DeclareAbbreviation\ibvs{Inf. Bull. Var. Stars}

\begin{document}
\SetRunningHead{T. Kato, H. Baba, and D. Nogami}{IR Com: Eclipsing Dwarf Nova
                Below the Period Gap}

\Received{}
\Accepted{}

\title{IR Com: Deeply Eclipsing Dwarf Nova Below the Period Gap \\
       --- A Twin of HT Cas?}

\author{Taichi \textsc{Kato}}
\affil{Department of Astronomy, Kyoto University,
       Sakyo-ku, Kyoto 606-8502}
\email{tkato@kusastro.kyoto-u.ac.jp}

\author{Hajime \textsc{Baba}}
\affil{Astronomical Data Analysis Center, National Astronomical
       Observatory, Mitaka, Tokyo 181-8588}
\email{hajime.baba@nao.ac.jp}

\email{\rm{and}}

\author{Daisaku \textsc{Nogami}}
\affil{Hida Observatory, Kyoto University, Kamitakara, Gifu 506-1314}
\email{nogami@kwasan.kyoto-u.ac.jp}



\maketitle

\begin{abstract}
   We observed an X-ray selected, deeply eclipsing cataclysmic variable
IR Com (=S 10932).  We detected an outburst occurring on 1996 January 1.
The light curve of the outburst closely resembled that of a normal
outburst of an SU UMa-type dwarf nova, rather than that of an intermediate
polar.  Time-resolved photometry during outburst showed that eclipses
became systematically deeper and narrower as the outburst faded.
Full-orbit light curves in quiescence showed little evidence of orbital
humps or asymmetry of eclipses.  In addition to the presence of high--low
transitions in quiescence, the overall behavior of outbursts and
characteristics of the eclipse profiles suggest that IR Com can be best
understood as a twin of HT Cas, a famous eclipsing SU UMa-type dwarf nova
with a number of peculiarities.
\end{abstract}

\section{Introduction}

   Cataclysmic variables (CVs) are close binary systems consisting of
a white dwarf and a red dwarf secondary transferring matter via the
Roche-lobe overflow.  Dwarf novae are a class of CVs showing outbursts,
which is believed to be a result of disk instabilities
[see \citet{osa96review} for a review].
Eclipsing CVs, especially eclipsing dwarf novae, provide a wonderful tool
for studying geometrically resolving structures of the accretion disk
(e.g. \cite{EclipseMapping}).

   IR Com (=S 10932) was first discovered as a ROSAT source,
RX J1239.5+2108 = 1RXS J123930.6+210815, which was identified with
a cataclysmic variable \citep{ric95ircom}.  \citet{ric95ircom} reported
that IR Com showed both high and low states, and occasional brightenings,
which resembled the behavior of a possible intermediate polar (IP), V426 Oph
\citep{wen90v426oph}.  \citet{wen95ircom} further revealed that IR Com is
an eclipsing CV with an orbital period of 0.08703 d.  The period is
just below the famous period gap in the distribution of orbital periods
of CVs, in which the number density of CVs is markedly reduced
(cf. \cite{RitterCV}).  Although \citet{wen95ircom} noted that most of CVs
with such a period are either SU UMa-type dwarf novae [for a recent review
of SU UMa-type stars and their observational properties, see
\citet{war95suuma}]
or polars, they rather regarded IR Com as an object of a possibly new
class, which shows high--low state transitions as well as infrequent dwarf
nova-like outbursts.

   We noticed the similarity of properties of IR Com with those of
HT Cas, a famous eclipsing SU UMa-type dwarf nova with a number of
peculiarities, and started observing IR Com since 1996.  On the very
first night of our observation (1996 January 1), we detected IR Com
in outburst \citep{kat96ircomalert306}.  The only known previous outbursts
were in 1959 February and 1988 February \citep{ric97ircom}.

\section{Observations}

\begin{figure}
  \begin{center}
    \FigureFile(88mm,60mm){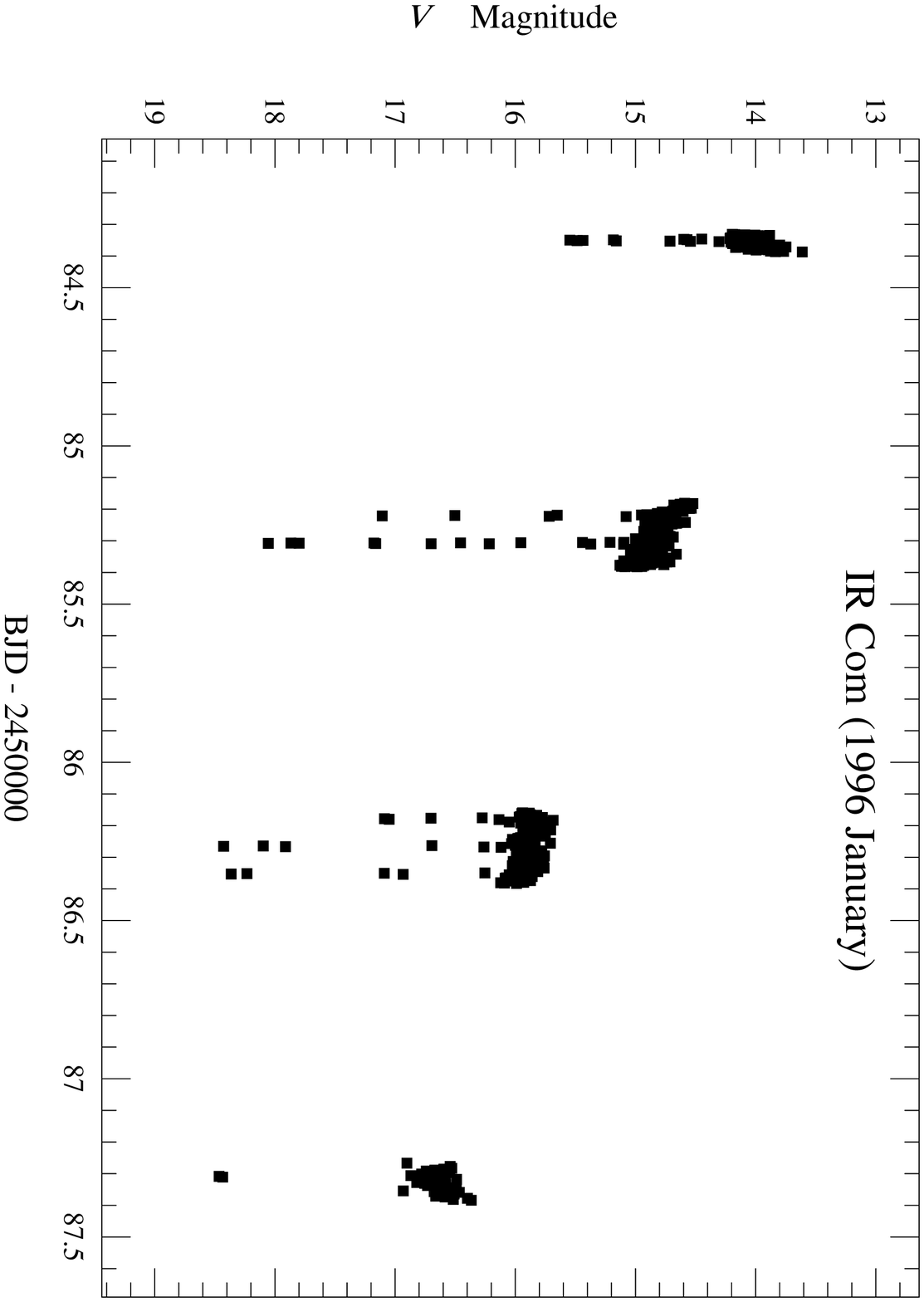}
  \end{center}
  \caption{The light curve of the 1996 January outburst.  Each point
  represents an average of 0.001 d bins.  Deep eclipses are superimposed
  on a fade at $\sim$0.9 mag d$^{-1}$.  Please note the averaging process
  slightly reduced the depths of eclipses (see figure \ref{fig:oeclph}
  more details of the eclipses).}
  \label{fig:burst}
\end{figure}

   The observations were carried out on 15 nights between 1996 January 1
and 1997 April 25, using a CCD camera (Thomson TH~7882, 576$\times$384
pixels) attached to the Cassegrain focus of the 60 cm reflector (focal
length=4.8 m) at Ouda Station, Kyoto University \citep{Ouda}.
An on-chip summation of 2$\times$2 pixels to one pixel was adopted.
An interference filter was used which had been designed to
reproduce the Johnson $V$ band.  The frames were first corrected for
standard de-biasing and flat fielding, and were then processed by
a microcomputer-based PSF photometry package developed by one of the
authors (TK).  The relative fluxes of the variable were determined
using GSC 1448.1951 ($V$=15.24) except for the first two runs.
The initial run on 1996 January 1 used GSC 1448.1216 ($V$=14.66), and the
first run on 1996 January 2 (BJD 2450085.180 -- .265; BJD = Barycentric
Julian Date) used BD +21$^{\circ}$ 2440 = GSC 1448.1508 ($V$=9.11) for
comparison stars.  The constancy of the comparison stars during the
observation was confirmed by using the common check star GSC 1448.2307
($V$=16.07).

   Barycentric corrections to the observed times were applied before the
following analysis.  The log of observations is summarized in table
\ref{tab:log}.

\begin{table*}
\caption{Log of observations.}\label{tab:log}
\begin{center}
\begin{tabular}{lccccc}
\hline\hline
Date      & BJD\commenta (start--end) & N\commentb & Mag\commentc &
            Error\commentd & Exp\commente \\
\hline
1996 Jan. 1  & 50084.331 -- 50084.387 &  68 & 14.00 & 0.02 & 60 \\
1996 Jan. 2  & 50085.180 -- 50085.265 & 463 & 14.71 & 0.01 & 7--9 \\
1996 Jan. 2  & 50085.270 -- 50085.382 & 256 & 14.87 & 0.01 & 30 \\
1996 Jan. 3  & 50086.159 -- 50086.383 & 402 & 15.89 & 0.01 & 40 \\
1996 Jan. 4  & 50087.265 -- 50087.386 & 130 & 16.62 & 0.01 & 60 \\
1996 Jan. 12 & 50095.225 -- 50095.333 &  91 & 16.91 & 0.03 & 90 \\
1996 Jan. 13 & 50096.184 -- 50096.344 &  89 & 16.93 & 0.03 & 90 \\
1996 Jan. 21 & 50104.272 -- 50104.338 &  48 & 17.23 & 0.02 & 90 \\
1996 Jan. 26 & 50109.229 -- 50109.279 &  42 & 16.88 & 0.02 & 90 \\
1996 Jan. 30 & 50113.216 -- 50113.271 &  23 & 17.17 & 0.04 & 90 \\
1996 Feb. 5  & 50119.302 -- 50119.381 &  21 & 17.10 & 0.08 & 90 \\
1996 Feb. 23 & 50137.102 -- 50137.186 &  60 & 17.09 & 0.02 & 90 \\
1996 Feb. 27 & 50141.086 -- 50141.195 &  27 & 17.52 & 0.12 & 90 \\
1996 Mar. 26 & 50169.114 -- 50169.136 &  49 & 15.36 & 0.02 & 30 \\
1997 Apr. 24 & 50563.148 -- 50563.246 & 115 & 17.18 & 0.02 & 60 \\
1997 Apr. 25 & 50564.158 -- 50564.254 & 112 & 17.56 & 0.02 & 60 \\
\hline
 \multicolumn{6}{l}{\commenta BJD$-$2400000.} \\
 \multicolumn{6}{l}{\commentb Number of frames.} \\
 \multicolumn{6}{l}{\commentc Averaged $V$ magnitude outside eclipses.} \\
 \multicolumn{6}{l}{\commentd Standard error of the averaged magnitude.} \\
 \multicolumn{6}{l}{\commente Exposure time (s).} \\
\end{tabular}
\end{center}
\end{table*}

\section{Results and discussion}

\subsection{Outburst observation}\label{sec:outburst}

   Although our observation did not cover the pre-maximum and maximum
phase of the 1996 January outburst, the object reached at least $V$=14.0,
and faded rapidly, reaching a rate of $\sim$0.9 mag d$^{-1}$ (figure
\ref{fig:burst}).  This decay from the outburst is quite characteristic
to that of normal outbursts of SU UMa-type dwarf novae, and unlike
brief (usually $<$1 d), rapidly fading ``outbursts" of intermediate
polars observed in TV Col \pasjcitep2sub{szk83tvcolflare}{szk84tvcolflare}
and EX Hya \pasjcitep2sub{hel89exhya}{hel00exhyaoutburst}.
The present observation seems to preclude the IP-type interpretation
of ``outbursts" of IR Com, that was originally proposed by
\citet{ric97ircom}.  No evidence of superhumps was detected.

   IR Com is now established as one of rare eclipsing dwarf novae which
show deep eclipses even during outbursts.  Only a very limited number of
such systems are known especially below the period gap (Z Cha, OY Car,
HT Cas, V2051 Oph, DV UMa and IY UMa).  Since most of dwarf novae below
the period gap are known to be SU UMa-type dwarf novae, the new specimen
of a deeply eclipsing dwarf nova below the period gap is expected to
provide an important role in studying the effect of the tidal instability
which is responsible for superoutbursts and superhumps.  Since the
mass-ratio ($q=M_2/M_1$) of this relatively long-period (0.08703 d)
system is expected to lie close to the stability border of the tidal
instability \citep{whi88tidal}, the role of IR Com would be especially
important in studying the tidal instability near the stability border.

\subsection{Eclipse ephemeris}\label{sec:ephem}

   Our observation has established that IR Com shows deep eclipses even
during outbursts (figure \ref{fig:burst}).  We have determined
mid-eclipse times by minimizing the dispersions of the eclipse light
curves folded at the mid-eclipse times.  The error of eclipse times were
estimated using the Lafler-Kinman class of methods, as applied by
\citet{fer89error}.  The validity of the estimated errors has been
confirmed by two independent methods: 1) application to different ranges
(in eclipse depth) of the data in order to test the effect of the
potential asymmetry, and 2) application to the binned data of the first
January 2 high-speed photometry run in order to test the effect of
reduced time resolution.  The tests have proven that the both effects
did not significantly affect the estimated errors\footnote{Readers, however,
  should bear in mind that \citet{ric97ircom} reported 25 s difference
  between their $B$ and $R$ observations.  Although $B$-band observations
  are expected to more susceptible to the spatial distribution of
  high-temperature structure in the disk (e.g. bright spot), our observation
  may have suffered from a lesser degree of similar systematic errors.
  The error estimates should therefore be treated as a statistical measure
  of the observational errors.
}.
Table \ref{tab:eclmin} summarizes the observed times of
eclipses (labeled as ``this work"), together with the published eclipse
times reported by \citet{ric97ircom}.  The times from \citet{ric97ircom}
have been converted into the BJD system, common to the present observation.
The cycle count ($E$) follows the definition by \citet{ric97ircom}.

   Eclipses before $E$=$-$24 having been chance detections on photographic
plates with long exposure times, we used eclipse times only after
BJD 2449484 in determining the eclipse ephemeris.  We first obtained
a linear regression to all the data.  After rejecting eclipse times having
$|O-C|>0.0005$ d ($E$=0, 3457, 3469, 6994), we obtained the following
linear ephemeris.  The orbital phases used in the following figures
and discussions are based on this equation.

\begin{equation}
\rm{BJD_{min}} = 2449486.48184(6) + 0.087038642(10) E. \label{equ:reg1}
\end{equation}

\begin{figure}
  \begin{center}
    \FigureFile(88mm,60mm){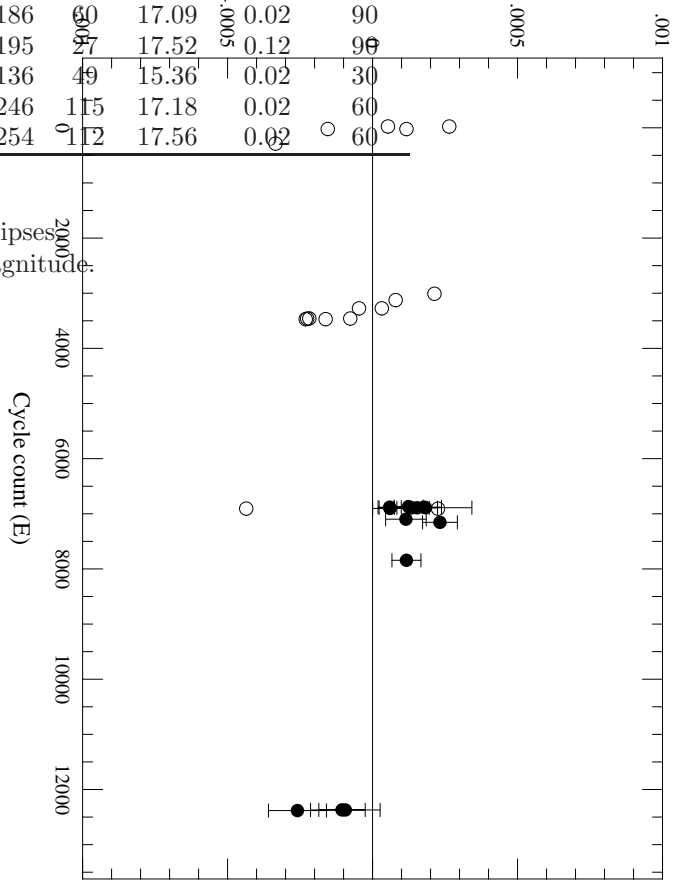}
  \end{center}
  \caption{$O-C$ diagram of eclipse minima.
  The $O-C$'s are calculated against equation \ref{equ:reg1}.  Filled
  circles with error bars and open circle represent this observation
  and \citet{ric97ircom}, respectively.  Although the $O-C$'s were rather
  constant between $E$=$-$24 and $E$=7843, there seems to be
  a systematic tendency of negative $O-C$'s after $E$=12370.}
  \label{fig:oc}
\end{figure}

   Figure \ref{fig:oc} shows the $O-C$ diagram of the eclipse centers
used to calculate equation \ref{equ:reg1}.  Although the $O-C$'s were rather
constant between $E$=$-$24 and $E$=7843, there seems to be a systematic
tendency of negative $O-C$'s after $E$=12370 (the deviation from the
linear ephemeris before $E$=12370 was $\sim$0.0003 d or $\sim$30 s)\footnote{
  The potential systematic $O-C$ variation, caused by different passbands
  and different outburst phases, is expected to be minimal, since the
  deviation was significant between the same $V$-band, outburst
  observations before and after $E$=12370.
}.
Fitting a quadratic equation to the observed times has only yielded
a marginally significant quadratic term of $-3.2\pm2.3 \times 10^{-12}
\times E^2$.

\begin{table}
\caption{Eclipses and $O-C$'s of IR Com.}\label{tab:eclmin}
\begin{center}
\begin{tabular}{lrrrc}
\hline\hline
Eclipse\commenta & Error\commentb & $E$\commentc & $O-C$\commentd
                 & Ref.\commente \\
\hline
37778.386   &  - & $-$134516 & $-$588 & 1 \\
45044.468   &  - & $-$51035  &  325   & 1 \\
45814.407   &  - & $-$42189  & $-$158 & 1 \\
46910.398   &  - & $-$29597  & $-$116 & 1 \\
47612.455   &  - & $-$21531  & $-$216 & 1 \\
49484.39297 &  - &   $-$24   &    5   & 1 \\
49484.48022 &  - &   $-$23   &   26   & 1 \\
49486.48240 &  - &       0   &   56   & 1 \\
49488.39654 &  - &      22   & $-$15  & 1 \\
49488.48385 &  - &      23   &   12   & 1 \\
49511.46160 &  - &     287   & $-$33  & 1 \\
49748.46837 &  - &    3010   &   21   & 1 \\
49758.47768 &  - &    3125   &    8   & 1 \\
49771.44639 &  - &    3274   &    3   & 1 \\
49771.53335 &  - &    3275   &  $-$5  & 1 \\
49787.37359 &  - &    3457   & $-$84  & 1 \\
49787.46125 &  - &    3458   & $-$22  & 1 \\
49787.54843 &  - &    3459   &  $-$8  & 1 \\
49787.63532 &  - &    3460   & $-$23  & 1 \\
49788.41973 &  - &    3469   &   84   & 1 \\
49788.50577 &  - &    3470   & $-$16  & 1 \\
49788.59274 &  - &    3471   & $-$23  & 1 \\
50084.35040 &  5 &    6869   &   12   & 2 \\
50085.22084 &  6 &    6879   &   18   & 2 \\
50085.30776 &  4 &    6880   &    6   & 2 \\
50086.17827 & 16 &    6890   &   18   & 2 \\
50086.26528 &  7 &    6891   &   15   & 2 \\
50086.35230 &  6 &    6892   &   14   & 2 \\
50087.30965 &  6 &    6903   &    6   & 2 \\
50087.57027 &  - &    6906   & $-$44  & 1 \\
50087.65797 &  - &    6907   &   23   & 1 \\
50095.23092 & 54 &    6994   &   81   & 2 \\
50104.28224 &  7 &    7098   &   12   & 2 \\
50109.24356 &  6 &    7155   &   23   & 2 \\
50169.12603 &  5 &    7843   &   12   & 2 \\
50563.14974 &  8 &   12370   & $-$11  & 2 \\
50563.23679 & 12 &   12371   &  $-$9  & 2 \\
50564.19405 & 10 &   12382   & $-$26  & 2 \\
\hline
 \multicolumn{5}{l}{\commenta Eclipse center.  BJD$-$2400000.} \\
 \multicolumn{5}{l}{\commentb Estimated error in 10$^{-5}$ d.} \\
 \multicolumn{5}{l}{\commentc Cycle count.} \\
 \multicolumn{5}{l}{\commentd Against equation \ref{equ:reg1}.
                           Unit in 10$^{-5}$ d.} \\
 \multicolumn{5}{l}{\commente Reference. 1: \citet{ric97ircom}, 2: this work} \\
\end{tabular}
\end{center}
\end{table}

\subsection{Eclipses in outburst}

\begin{figure}
  \begin{center}
    \FigureFile(88mm,120mm){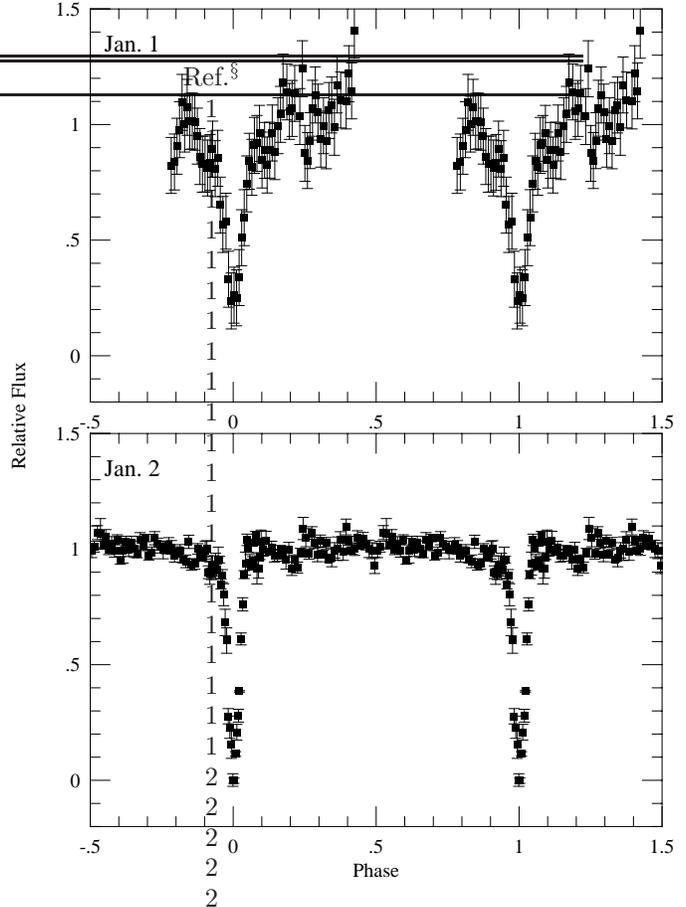}
  \end{center}
  \caption{Phase-averaged light curves during the 1996 January
  outburst.  Each point represents an average and standard error of
  each 0.01 phase bin.
  The phases are calculated against equation \ref{equ:reg1}.
  The fluxes are normalized to 1 outside eclipses.  The trend of linear
  decline from the outburst maximum was subtracted from the January 2
  data.}
  \label{fig:oecl1}
\end{figure}

\begin{figure}
  \begin{center}
    \FigureFile(88mm,120mm){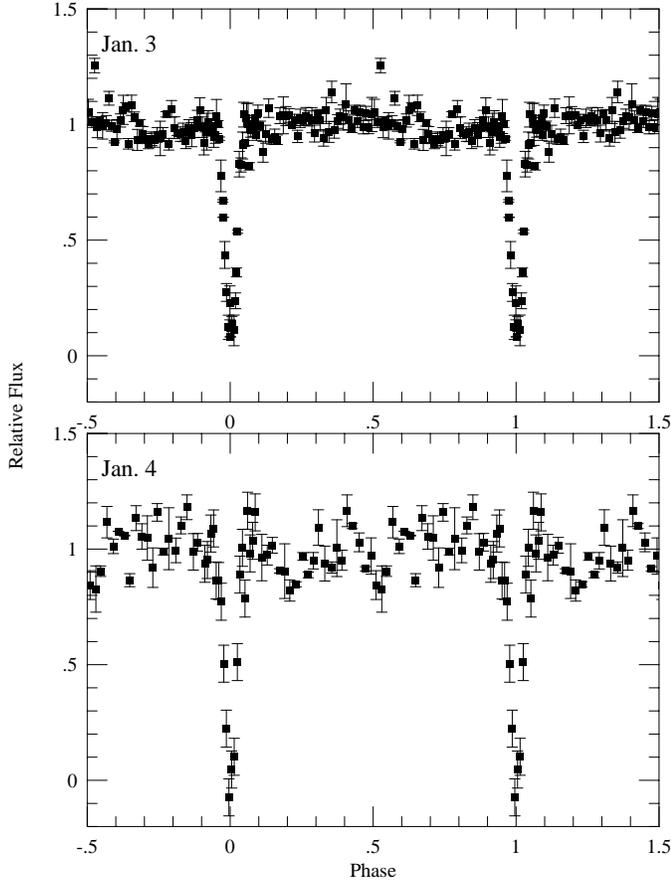}
  \end{center}
  \caption{Phase-averaged light curves during the 1996 January
  outburst (continued).  Each point represents an average and standard error
  of each 0.01 phase bin.  The fluxes are normalized to 1 outside eclipses.
  The trend of linear decline from the outburst maximum was subtracted.}
  \label{fig:oecl2}
\end{figure}

   Figures \ref{fig:oecl1} and \ref{fig:oecl2} show phase-averaged light
curves during the 1996 January outburst, obtained on four successive
nights.  The phases are calculated against equation \ref{equ:reg1}.
The trend of linear decline from the outburst maximum was first subtracted
from the data, using a linear fit to the non-eclipsed portion of the light
curve.  No subtraction of the linear trend was made for the data on
January 1, which showed a slight tendency of brightening, but the
observation was too short to meaningfully determine the rate.
The fluxes were then normalized to 1 outside eclipses.
Orbital humps outside eclipses were not evident.

\begin{figure}
  \begin{center}
    \FigureFile(88mm,120mm){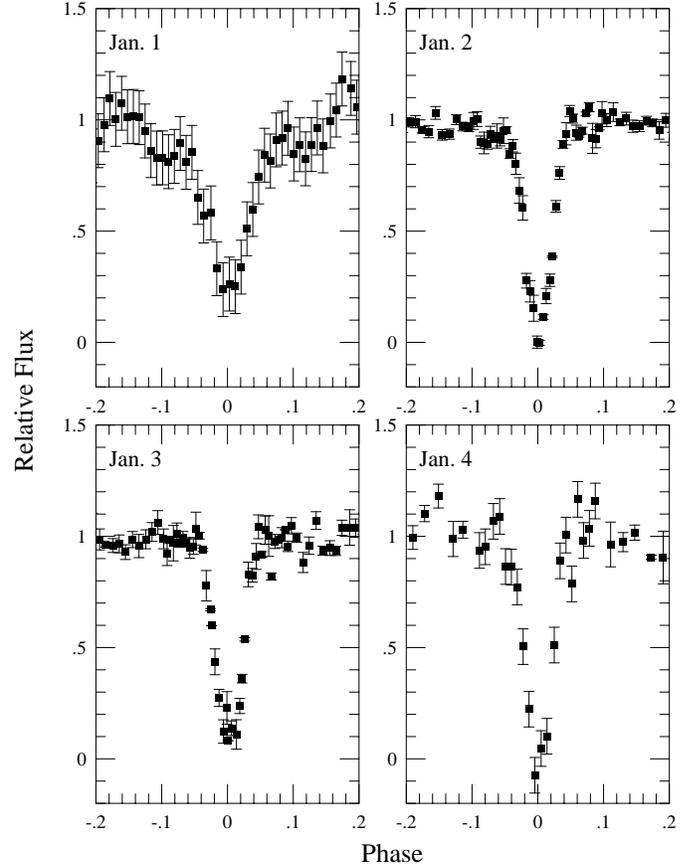}
  \end{center}
  \caption{Enlarged light curves of eclipses during the 1996 January
  outburst.  Each point represents an average and standard error of
  each 0.01 phase bin.  The fluxes are normalized to 1 outside eclipses.
  The trend of linear decline from the outburst maximum was subtracted
  except for the January 1 data.}
  \label{fig:oeclph}
\end{figure}

   Figure \ref{fig:oeclph} shows the enlargement of nightly eclipse
profiles during the outburst.  The eclipse on January 1 (presumably close
to the outburst maximum, or even just before the maximum) had a depth of
1.5$\pm$0.2 mag, and a full width of 0.14 in phase.  There also seem to be
a broader, shallow fading between phases $-$0.13 and 0.15, although the
origin of this feature is unknown.  Eclipses became narrower and deeper
as the object faded (figure \ref{fig:oeclph}; table \ref{tab:eclpar}),
indicating that the luminous part of the accretion disk became smaller.
This feature closely resembles the change in the eclipse profile observed
during normal outbursts of other dwarf novae (e.g. HT Cas: \cite{ioa99htcas};
\cite{bab99htcas}; Baba et al., in preparation).

\begin{table}
\caption{Variation of the eclipse width and depth.}\label{tab:eclpar}
\begin{center}
\begin{tabular}{lcc}
\hline\hline
Date        & Width (phase)\commenta & Depth (mag)\commentb \\
\hline
1996 Jan. 1 & 0.14 & 1.5 \\
1996 Jan. 2 & 0.09 & 2.3 \\
1996 Jan. 3 & 0.08 & 2.4 \\
1996 Jan. 4\commentc & 0.09 & 2.8 \\
\hline
 \multicolumn{3}{l}{\commenta Uncertainty 0.01.} \\
 \multicolumn{3}{l}{\commentb Uncertainty 0.2.} \\
 \multicolumn{3}{l}{\commentc Slightly uncertain values because of the low} \\
 \multicolumn{3}{l}{\phantom{\commentc} signal-to-noise ratio.} \\
\end{tabular}
\end{center}
\end{table}

\begin{figure}
  \begin{center}
    \FigureFile(88mm,60mm){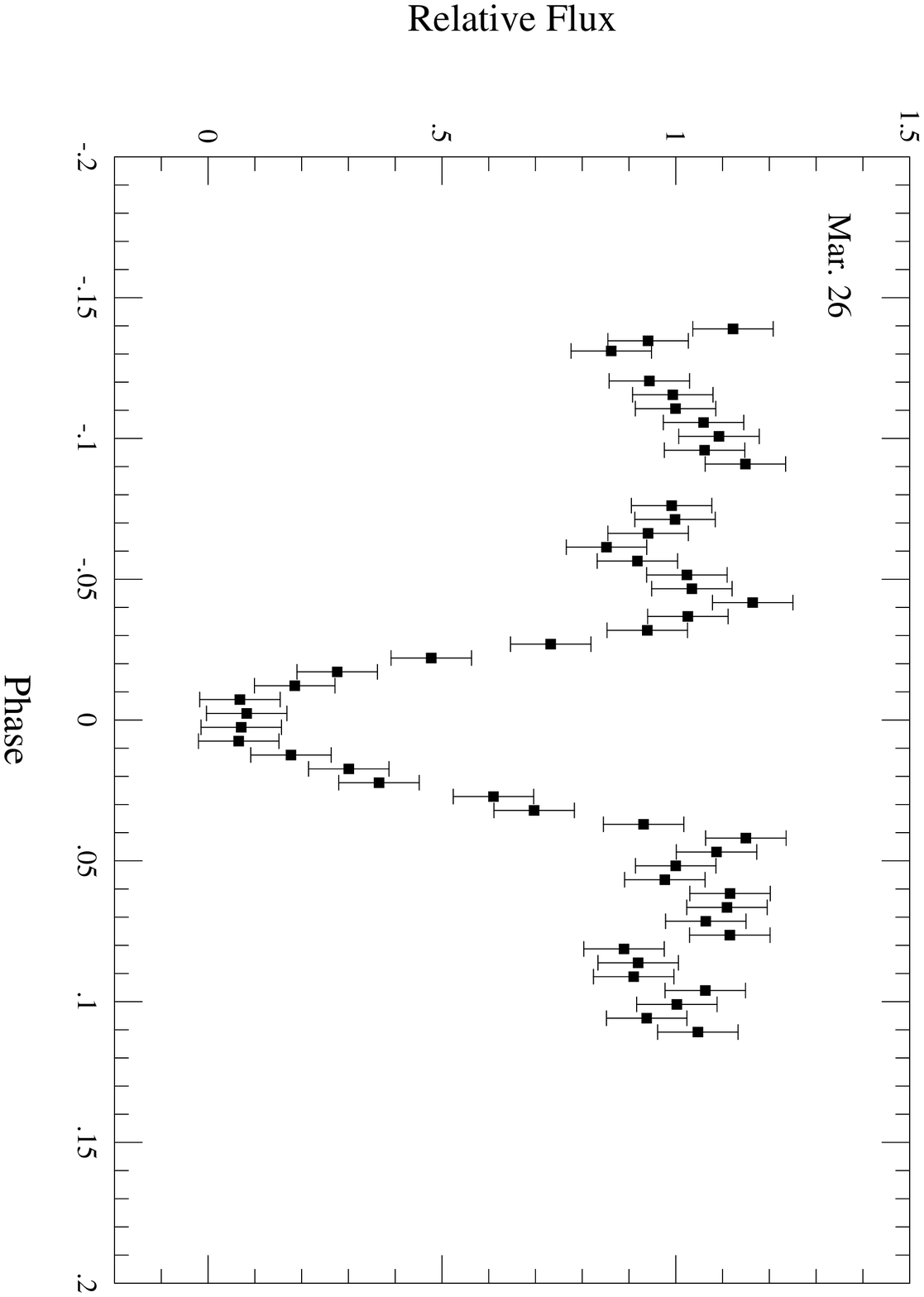}
  \end{center}
  \caption{Eclipse light curve on 1996 March 26 (in outburst).
  Each point represents an average and standard error of each 0.01 phase bin.
  The fluxes are normalized to 1 outside eclipses.}
  \label{fig:oeclmar}
\end{figure}

   We also obtained time-resolved photometry during the 1996 March
outburst, upon the alert by M. Iida (Variable Star Observers League
in Japan).  The result shown in figure \ref{fig:oeclmar} is essentially
same as observed in the 1996 January outburst.  Since the observed eclipses
were deeper than those observed during the early stage of the 1996 January
outburst, the outburst should have been caught during the later stage of
an outburst.

\subsection{Quiescent eclipses}\label{sec:quiecl}

\begin{figure}
  \begin{center}
    \FigureFile(80mm,110mm){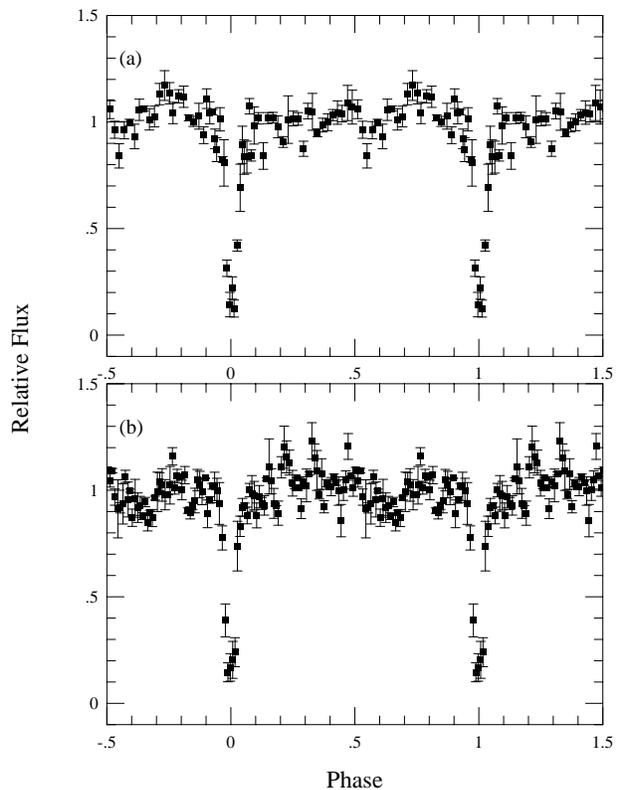}
  \end{center}
  \caption{Phase-averaged quiescent light curves on two epochs.
  Each point represents an average and standard error of each 0.017 (panel
  a) or 0.01 (panel b) phase bin.  The phases are calculated against
  equation \ref{equ:reg1}.
  The fluxes are normalized to 1 outside eclipses.
  The panel (a) represents the quiescent period between
  1996 January 12 and 26 (BJD 2450095.2 -- 2450109.3).
  The panel (b) represents the quiescent period between
  1997 April 24 and 25 (BJD 2450563.1 -- 2450564.3).  Deep, narrow
  eclipses are evident.  Orbital humps are very weak or almost absent.}
  \label{fig:quecl}
\end{figure}

   Figure \ref{fig:quecl} shows phase-averaged light curves on two
quiescent epochs.  The phases are calculated against equation \ref{equ:reg1}.
The fluxes are normalized to 1 outside eclipses.  A slow fading trend
on 1997 April 24 was removed by a linear fit to the observations outside
eclipses before normalization.  The panel (a) represents the period between
1996 January 12 and 26 (BJD 2450095.2 -- 2450109.3).  The panel (b)
represents the period between 1997 April 24 and 25 (BJD 2450563.1 --
2450564.3).  On both panels, deep (2.2$\pm$0.2 and 2.1$\pm$0.2 mag,
respectively) and narrow (full widths 0.13 and 0.12 in phase,
respectively) eclipses are evident.  The profile of the eclipse and
full-orbit light curve very much resembles that presented by
\citet{ric97ircom}.
The most remarkable feature is the very weak (panel a), or almost absent
(panel b) orbital humps preceding eclipses.  Such a feature is rarely
observed in high-inclination dwarf novae (see subsection \ref{sec:comp:hump}
for a further discussion).

\subsection{$O-C$ change}

   As seen in figure \ref{fig:oc}, the eclipse timings were well
represented by a constant period for the period 1994--1996.  Looking more
closely at the 1996 data, which covered outburst and quiescent stages,
there was no significant difference of $O-C$'s between different brightness
phases, in outburst and in quiescence.  In most deeply eclipsing dwarf novae,
the asymmetry of the eclipse light curve in quiescence (usually produced
by the presence of a bright spot) results in a significant offset of
eclipse centers in quiescence against the eclipse center of the white
dwarf, or against eclipse centers observed in outburst.  The apparent
lack of this effect in IR Com suggests that the asymmetry of the accretion
disk in quiescence is weak, which is in good agreement with the absence
of orbital humps in quiescence (subsection \ref{sec:quiecl}).

   $O-C$'s showed a small, but a statistically significant change between
1996 and 1997 (subsection \ref{sec:ephem}).
Fitting a quadratic equation to all observed
times between 1994 and 1997 has yielded a marginal quadratic term of
$-3.2\pm2.3 \times 10^{-12} \times E^2$.  This value may show the same
order of a secular period change suggested in an eclipsing dwarf nova
Z Cha ($\dot{P}$=+1.6$\pm$0.2 $\times$ 10$^{-12}$ cycle$^{-1}$:
\cite{coo81zchapdot}).
The possibility of a secular period change, however, should be confirmed
by further observations, since this may represent a more sporadic or
quasi-cyclic $O-C$ change as reported in other eclipsing dwarf novae
(e.g. IP Peg: \cite{woo89ippeg}).

\section{Comparison with HT Cas}

   Among the above eclipsing dwarf novae below the period gap, HT Cas
is known to have peculiar characteristics.   They can be summarized as
below.

\begin{enumerate}

\item\label{item:cycle}
   Among deeply eclipsing SU UMa-type dwarf novae, HT Cas does not
   show a clear supercycle (a cycle between successive superoutbursts).
   \citet{wen87htcas} studied Sonneberg plates and obtained a mean cycle
   length of 400$\pm$50 d.  However, no superoutburst has been observed
   up to 2001, since the last one in 1985.  The observed frequency of
   outbursts is lower than that expected from the observed quiescence
   mass-transfer rate ($\dot{M}$) (\cite{zha86htcas}; \cite{woo95htcasXray}).

\item\label{item:states}
   Both high (bright) and low (faint) states exist in quiescence.
   High states are typically $\sim$1 mag brighter than low states
   (\cite{zha86htcas}; \cite{woo95htcasXray}; \cite{rob96htcas}).

\item\label{item:hump}
   Orbital humps, which are considered to reflect the bright spot
   of the accretion impact point on the disk, are not prominent, and
   only occasionally seen in high-state quiescence (\cite{zha86htcas};
   \cite{pat81DNOhtcas}; \cite{woo95htcasXray}).  The profile of eclipses
   strongly varies.  The orbital humps are absent in low-state quiescence
   [a representative collection of low-state orbital light curves can be
   seen in \citet{woo95htcasXray}, which also presents an example of the
   presence of orbital humps in high-state quiescence].
   Eclipse mapping of flickering shows a strong concentration toward the
   inner disk
   (\cite{wel95htcas}; \cite{bru00htcasv2051ophippeguxumaflickering}),
   in contrast to the classical example of U Gem, whose flickering is
   known to be strongly concentrated in the bright spot \citep{war71ugem}.

\item\label{item:xray}
   HT Cas shows moderately strong X-ray emission relative to the optical
   flux ($f_{\rm X}/f_{\rm opt}$) among non-magnetic dwarf novae
   (\cite{ver97ROSAT}; see also \cite{woo95htcasXray} and
   \cite{muk97htcas} for the detailed analysis of ROSAT pointed
   observation).

\end{enumerate}

   IR Com exhibits a number of similarities with HT Cas.  We examine
them in more detail in the following subsections.

\subsection{Outburst cycle length}

   \citet{wen95ircom} and \citet{ric97ircom} suggested that the
recurrence time of major outbursts is 8/$N$ yr, where $N$($>0$) is
a small integer.  Although more recent observations by the VSNET
Collaboration\footnote{
  $\langle$http://www.kusastro.kyoto-u.ac.jp/vsnet/$\rangle$}
suggest a higher frequency of outbursts, only five outbursts (table
\ref{tab:outbursts}) are known between 1996--2001, in spite of intensive
monitoring.  All of them were short outbursts, lasting less than a few days.
No clear periodicity can be found from these data, supporting the
finding by \citet{ric97ircom}.  This irregular, infrequent occurrence of
short outbursts (normal outbursts) is very reminiscent of the irregular
behavior of HT Cas \citep{wen87htcas}.  The 1988 outburst reaching mag 13.5
reported by \citet{ric97ircom} lasted at least two days, and showed
a relatively slow rise.  Although the available information is very
limited to draw a firm conclusion, this outburst may have been
a superoutburst.
The overall characteristics of outbursts of IR Com closely resembles those
of HT Cas (item \ref{item:cycle}).

\begin{table}
\caption{Outbursts of IR Com between 1996--2001.}\label{tab:outbursts}
\begin{center}
\begin{tabular}{lcc}
\hline\hline
Date         &  JD     & peak magnitude \\
\hline
1996 Jan. 1  & 2450084 & 14.0 \\
1996 Mar. 26 & 2450169 & 14.7 \\
1997 Apr. 23 & 2450562 & 14.1 \\
1998 Jun. 27 & 2450991 & 14.3 \\
2001 Mar. 25 & 2451993 & 14.0 \\
\hline
\end{tabular}
\end{center}
\end{table}

\subsection{High and low states in quiescence}\label{sec:comp:state}

   IR Com is known to show both high and low states in quiescence
(\cite{ric95ircom}; \cite{ric97ircom}).  The quiescent observation of the
present work was mainly done in high state (at $V$=16.9--17.2).  However,
the system was reported to show an excursion to a low state ($V\sim$19)
in 1996 June \citep{kro96ircomalert428}, only two month after our final
observation in 1996.  The intermediately faint observation ($V$=17.5)
on 1996 February 27 may be a suggestion of an ongoing excursion to a faint
state.
The existence of high and low states in quiescence, and the time-scales
of transitions between them (a hundred to several hundreds days), are
similar to those \citep{rob96htcas} observed in HT Cas
(item \ref{item:states}).

\subsection{Orbital humps}\label{sec:comp:hump}

   As shown in \ref{sec:quiecl}, we did not detect significant
orbital humps during our quiescent observations.  This is unusual for
an eclipsing dwarf nova, and more resembles the quiescence of HT Cas.
Since most of our observations were done during high-state
quiescence of IR Com (subsection \ref{sec:comp:state}), further
observations of eclipses and humps in IR Com in low quiescent state are
therefore highly wanted in order to establish the similarity with HT Cas.

\subsection{X-ray observations}

   As originally selected in X-ray surveys, IR Com emits relatively
strong (and likely hard) X-rays \citep{ric95ircom}.  Table \ref{tab:xray}
shows the comparison of IR Com with HT Cas.  Although direct comparison is
difficult, because of different interstellar absorption and poorly
determined X-ray spectrum in IR Com, it is evident both systems have
a very similar $f_{\rm X}/f_{\rm opt}$, which is higher than the average
dwarf novae.  Both systems are apparently non-magnetic, from the absence
of coherent modulations either in X-rays or optical (\cite{ric95ircom}
for IR Com).  Since most of the known X-ray luminous CVs are magnetic CVs
\citep{ver97ROSAT}, the high $f_{\rm X}/f_{\rm opt}$ in IR Com,
an apparently non-magnetic dwarf nova, may require a similar explanation
to HT Cas to effectively produce X-ray emissions from the boundary layer
(e.g. \cite{woo95htcasXray}; \cite{muk97htcas}).
Future phase-resolved X-ray observations of IR Com will be valuable in
testing the similarity of processes of X-ray emission, as well as the
accretion processes in the quiescent disk, between IR Com and HT Cas.

\begin{table}
\caption{Comparison of X-ray properties between IR Com and HT Cas from
the ROSAT 1RXS Catalogue.}\label{tab:xray}
\begin{center}
\begin{tabular}{cccc}
\hline\hline
Object & ctr\commenta & $V$ mag\commentb & $f_X$/$f_{\rm opt}$\commentc \\
\hline
IR Com &   0.061 & 17.0 & $-$0.25 \\
HT Cas &   0.099 & 16.4 & $-$0.07 \\
\hline
 \multicolumn{4}{l}{\commenta Total count rate (s$^{-1}$).} \\
 \multicolumn{4}{l}{\commentb Typical quiescent $V$ magnitude.} \\
 \multicolumn{4}{l}{\commentc Taken from \citet{ROSATRXP}, electronic} \\
 \multicolumn{4}{l}{\phantom{\commentc} catalogue table cor\_pri.dat.} \\
\end{tabular}
\end{center}
\end{table}

\section{Conclusion}

   We observed an X-ray selected, deeply eclipsing cataclysmic variable
IR Com (=S 10932) in outburst and quiescence.  The light curve of the
outburst, which occurred on 1996 January 1, was indistinguishable from
that of a normal outburst of an SU UMa-type dwarf nova.  Time-resolved
photometry during outburst showed that the evolution of the eclipse light
curve is a typical one for a dwarf nova outburst.  Full-orbit
light curves in quiescence show little evidence of orbital humps or
asymmetry of eclipses.  In addition to the presence of high--low
transitions in quiescence, the overall behavior of outbursts and
characteristics of the eclipse profiles suggest that IR Com can be
best understood as a twin of HT Cas, an eclipsing SU UMa-type dwarf
nova with a number of peculiarities.

\vskip 3mm

   The authors are grateful to VSNET members for providing vital
observations, and to Mr. M. Iida for promptly notifying us of the 1996
March outburst.


\begin{thebibliography}{}

\bibitem[Baba et~al.\labelspace(1999)]{bab99htcas}
  Baba, H., Kato, T., Nogami, D., \& Hirata, R.\ 1999, in Disk Instabilities in
  Close Binary Systems, ed. S. Mineshige, \& J.~C. Wheeler (Universal Academy
  Press, Tokyo), p.~123

\bibitem[Bruch\labelspace(2000)]{bru00htcasv2051ophippeguxumaflickering}
  Bruch, A.\ 2000, \aap, 359, 998

\bibitem[Cook, Warner\labelspace(1981)]{coo81zchapdot}
  Cook, M.~C., \& Warner, B.\ 1981, \mnras, 196, 55P

\bibitem[Fernie\labelspace(1989)]{fer89error}
  Fernie, J.~D.\ 1989, \pasp, 101, 225

\bibitem[Hellier et~al.\labelspace(2000)]{hel00exhyaoutburst}
  Hellier, C., Kemp, J., Naylor, T., Bateson, F.~M., Jones, A., Overbeek, D.,
  Stubbings, R., \& Mukai, K.\ 2000, \mnras, 313, 703

\bibitem[Hellier et~al.\labelspace(1989)]{hel89exhya}
  Hellier, C., Mason, K.~O., Smale, A.~P., Corbet, R. H.~D., O'Donoghue, D.,
  Barrett, P.~E., \& Warner, B.\ 1989, \mnras, 238, 1107

\bibitem[Horne\labelspace(1985)]{EclipseMapping}
  Horne, K.\ 1985, \mnras, 213, 129

\bibitem[Ioannou et~al.\labelspace(1999)]{ioa99htcas}
  Ioannou, Z., Naylor, T., Welsh, W.~F., Catal\'{a}n, M.~S., Worraker, W.~J.,
  \& James, N.~D.\ 1999, \mnras, 310, 398

\bibitem[Kato\labelspace(1996)]{kat96ircomalert306}
  Kato, T.\ 1996, \vsnetalert{306} \\
  http://www.kusastro.kyoto-u.ac.jp/vsnet/Mail/alert0/msg00306.html

\bibitem[Kroll\labelspace(1996)]{kro96ircomalert428}
  Kroll, P.\ 1996, \vsnetalert{428} \\
  http://www.kusastro.kyoto-u.ac.jp/vsnet/Mail/alert0/msg00428.html

\bibitem[Mukai et~al.\labelspace(1997)]{muk97htcas}
  Mukai, K., Wood, J.~H., Naylor, T., Schlegel, E.~M., \& Swank, J.~H.\ 1997,
  \apj, 475, 812

\bibitem[Ohtani et~al.\labelspace(1992)]{Ouda}
  Ohtani, H., Uesugi, A., Tomita, Y., Yoshida, M., Kosugi, G., Noumaru, J.,
  Araya, S., \& Ohta, K.\ 1992, Memoirs of the Faculty of Science, Kyoto
  University, Series A of Physics, Astrophysics, Geophysics and Chemistry, 38,
  167

\bibitem[Osaki\labelspace(1996)]{osa96review}
  Osaki, Y.\ 1996, \pasp, 108, 39

\bibitem[Patterson\labelspace(1981)]{pat81DNOhtcas}
  Patterson, J.\ 1981, \apjs, 45, 517

\bibitem[Richter, Greiner\labelspace(1995)]{ric95ircom}
  Richter, G.~A., \& Greiner, J.\ 1995, in Cataclysmic Variables, ed. A.
  Bianchini, M. Della~Valle, \& M. Orio (Kluwer Academic Publishers,
  Dordrecht), p.~177

\bibitem[Richter et~al.\labelspace(1997)]{ric97ircom}
  Richter, G.~A., Kroll, P., Greiner, J., Wenzel, W., Luthardt, R., \&
  Schwartz, R.\ 1997, \aap, 325, 994

\bibitem[Ritter, Kolb\labelspace(1998)]{RitterCV}
  Ritter, H., \& Kolb, U.\ 1998, \aaps, 129, 83

\bibitem[Robertson, Honeycutt\labelspace(1996)]{rob96htcas}
  Robertson, J.~W., \& Honeycutt, R.~K.\ 1996, \aj, 112, 2248

\bibitem[Szkody, Mateo\labelspace(1983)]{szk83tvcolflare}
  Szkody, P., \& Mateo, M.\ 1983, \pasp, 95, 596

\bibitem[Szkody, Mateo\labelspace(1984)]{szk84tvcolflare}
  Szkody, P., \& Mateo, M.\ 1984, \apj, 280, 729

\bibitem[Verbunt et~al.\labelspace(1997)]{ver97ROSAT}
  Verbunt, F., Bunk, W.~H., Ritter, H., \& Pfeffermann, E.\ 1997, \aap, 327,
  602

\bibitem[Voges et~al.\labelspace(1999)]{ROSATRXP}
  Voges, W., Aschenbach, B., Boller, T., Braeuninger, H., Briel, U., Burkert,
  W., Dennerl, K., Englhauser, J., Gruber, R., Haberl, F., Hartner, G.,
  Hasinger, G., Kuerster, M., Pfeffermann, E., Pietsch, W., Predehl, P., Rosso,
  C., M., S. J. H.~M., Truemper, J., \& Zimmermann, H.~U.\ 1999, \aap, 349, 389

\bibitem[Warner\labelspace(1995)]{war95suuma}
  Warner, B.\ 1995, \apss, 226, 187

\bibitem[Warner, Nather\labelspace(1971)]{war71ugem}
  Warner, B., \& Nather, R.~E.\ 1971, \mnras, 152, 219

\bibitem[Welsh, Wood\labelspace(1995)]{wel95htcas}
  Welsh, W., \& Wood, J.~H.\ 1995, in Flares and Flashes, ed. J. Greiner, H.~W.
  Duerbeck, \& R.~E. Gershberg (Springer-Verlag), p.~300

\bibitem[Wenzel\labelspace(1987)]{wen87htcas}
  Wenzel, W.\ 1987, \an, 308, 75

\bibitem[Wenzel et~al.\labelspace(1995)]{wen95ircom}
  Wenzel, W., Richter, G.~A., Luthardt, R., \& Schwartz, R.\ 1995, \ibvs, 4182

\bibitem[Wenzel, Splittgerber\labelspace(1990)]{wen90v426oph}
  Wenzel, W., \& Splittgerber, E.\ 1990, \ibvs, 3532

\bibitem[Whitehurst\labelspace(1988)]{whi88tidal}
  Whitehurst, R.\ 1988, \mnras, 232, 35

\bibitem[Wood et~al.\labelspace(1989)]{woo89ippeg}
  Wood, J.~H., Marsh, T.~R., Robinson, E.~L., Stiening, R.~F., Horne, K.,
  Stover, R.~J., Schoembs, R., Allen, S.~L., Bond, H.~E., Jones, D. H.~P.,
  Grauer, A.~D., \& Ciardullo, R.\ 1989, \mnras, 239, 809

\bibitem[Wood et~al.\labelspace(1995)]{woo95htcasXray}
  Wood, J.~H., Naylor, T., Hassall, B. J.~M., \& F., R.~T.\ 1995, \mnras, 273,
  772

\bibitem[Zhang et~al.\labelspace(1986)]{zha86htcas}
  Zhang, E.~H., Robinson, E.~L., \& Nather, R.~E.\ 1986, \apj, 305, 740

\end{thebibliography}
\end{document}